\documentclass[aps,onecolumn,nofootinbib,superscriptaddress]{revtex4}
\usepackage{amsmath,amssymb,bm}
\usepackage{graphicx}
\usepackage{epstopdf}
\usepackage{latexsym}
\usepackage{subfigure}
\usepackage{color}

\begin{document}
\title{Interplay of Quantum Criticality and Geometric Frustration in Columbite}
\author{SungBin Lee}
\affiliation{Department of Physics, University of California, Santa Barbara, CA-93106-9530}
\author{Ribhu K. Kaul\footnote{On leave of absence from:
     Department of Physics, University of Kentucky, Lexington, KY- 40506-0055}}
\affiliation{Microsoft Station Q, University of California, Santa Barbara, CA-93106-6105}
\author{Leon Balents}
\affiliation{Kavli Institute for Theoretical Physics, University of California, Santa Barbara, CA-93106-9530}
\date{\today}
\begin{abstract}
  Motivated by CoNb$_2$O$_6$ (belonging to the columbite family of
  minerals), we theoretically study the physics of quantum ferromagnetic
  Ising chains coupled anti-ferromagnetically on a triangular lattice in
  the plane perpendicular to the chain direction. We combine exact
  solutions of the chain physics with perturbative approximations for
  the transverse couplings. When the triangular lattice has an isosceles
  distortion (which occurs in the real material), the $T=0$ phase
  diagram is rich with five different states of matter: ferrimagnetic,
  N\'eel, anti-ferromagnetic, paramagnetic and incommensurate phases,
  separated by quantum phase transitions. Implications of our results to
  experiments on CoNb$_2$O$_6$ are discussed.
\end{abstract}

\maketitle

Two of the most exciting themes in condensed matter physics are the
destruction of magnetic order at absolute zero temperature, {\em i.e.}
quantum criticality~\cite{subir}, and quantum fluctuations in
geometrically frustrated magnets.   Each of these themes has its
archetype: the quantum Ising chain in a
transverse field~\cite{subir} for quantum criticality, and the triangular
lattice Ising model for geometrical frustration 
~\cite{Wannier}.   Remarkably both these ideals are
combined in CoNb$_2$O$_6$, a
material which may be understood as a collection of weakly coupled Ising
chains that sit on a triangular lattice in the plane perpendicular to
the chain direction. Very recently, CoNb$_2$O$_6$ has been the subject
of beautiful single crystal inelastic neutron scattering experiments by
Coldea {\em et al}~\cite{coldeaXX}. The observed spectrum has been shown
to reflect some very subtle properties of one-dimensional Ising quantum
critical physics, that were discovered three decades after the study of the
quantum Ising chain was initiated~\cite{lsm61}, in a triumph of theoretical physics
led by Zamolodchikov~\cite{zamo89}.   A full interpretation of the
experimental data requires an understanding of the interplay of
quantum criticality in the chain and the physics of frustration in the
plane perpendicular to the chain. In this paper, we combine the
remarkable insights of Zamolodchikov with the novelty that frustration
brings to the phase diagram of this remarkable material, setting up a
platform to understand the experimental data in totality. Our focus here
is to chart out the global phase diagram of CoNb$_2$O$_6$, as well as
other less well studied materials that have similar structures.

CoNb$_2$O$_6$ is an insulator. A strong single ion anisotropy forces the
spin on the magnetically active Co$^{2+}$ (3d$^7$) ions to point either
``up'' or ``down'' along the local easy axis, realizing a two-level
system. The Co$^{2+}$ ions interact with each other through Ising
interactions of the form $S^z_i S^z_j$, where $\vec{S}_i$ are the usual
spin-1/2 operators acting on the two level system located at site $i$. A
magnetic field applied along the b-axis is transverse to all the easy
axes and hence may be modeled by a term, $h_\perp S^x_i$, which acts
like a transverse field on the Ising spins. In the absence of the
transverse field the exchange interactions will cause the Ising spins to
order magnetically. On the other hand when $h_\perp$ is much larger than
the exchange interactions, the Ising spins will all lie in the quantum
superposition, $(|\uparrow \rangle + |\downarrow \rangle )/\sqrt{2}$,
highlighting the fact that the transverse field is a quantum mechanical
parameter with no classical analogue. The central question of interest
that we will address in this publication is: What is the nature of the
magnetic ordering and what is the sequence of $T=0$ phases between the
two extreme limiting cases as $h_\perp$ is tuned?

\begin{figure}
  \includegraphics[width=3in]{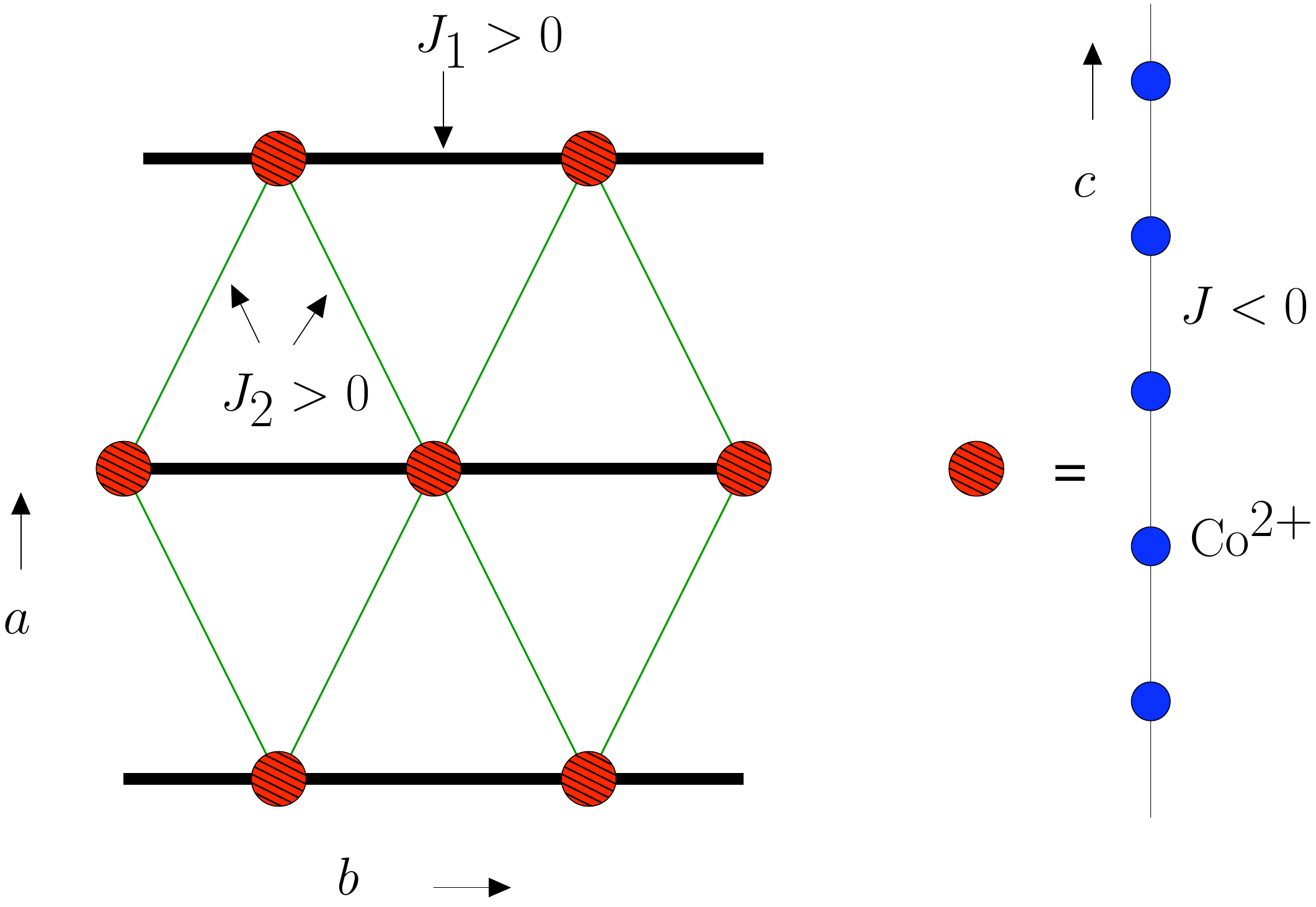}
  \caption{ \label{fig:isosceles} Schematic structure of Co$^{2+}$ ions
    in CoNb$_2$O$_6$.  Ising spins localized on each Co$^{2+}$ form
    ferromagnetic chains running along the c-direction; $J_0$ is the
    strength of the exchange along these c-axis chains. In the basal a-b
    plane perpendicular to the c-axis, the chains form a triangular
    lattice.  The stronger anti-ferro $J_1$ bonds parallel to the b-axis
    divide the basal triangular lattice into ``basal chains'' (marked
    with thick black lines), the weaker anti-ferro $J_2 = \alpha J_1$
    bonds (marked with thin green lines) connect neighboring ``basal
    chains''. $1-\alpha\ll 1$ in CoNb$_2$O$_6$ is a measure of the
    departure from a perfect basal triangular lattice in this material.
  }

\end{figure}

We begin by presenting a minimal theoretical model for CoNb$_2$O$_6$
which incorporates details of the columbite structure and which we
expect to reproduce most of the features of the experiments.  The
lattice consists of zig-zag chains of Co$^{2+}$ ions with a strong
ferromagnetic Ising interaction, $J_0<0$ along the c-direction. The
chains form an isosceles triangular lattice in the a-b plane, with
anti-ferromagnetic coupling to neighbors of strength $J_1$ and $J_2$ as
shown in Fig.~\ref{fig:isosceles}.
\begin{equation}
H = J_0 \sum S^z_{z; {\bf r}}S^z_{z+1; {\bf r}} 
+ J_1\sum_{{\bf r}} S^z_{z;{\bf r}}S^z_{z;{\bf r+a_1}} 
+ J_2\sum_{{\bf r}} S^z_{z;{\bf r}}S^z_{z;{\bf r+a_2}} 
+ J_2\sum_{{\bf r}} S^z_{z;{\bf r}}S^z_{z;{\bf r+a_3}} 
- h_{\perp} \sum
S^x_{z;{\bf r}},
\label{eq:model}
\end{equation}
${\bf a_{1,2,3}}$ are the three nearest neighbors on the triangular lattice illustrated
in Fig.~\ref{fig:isosceles}. Motivated by CoNb$_2$O$_6$
we focus on the limit when $0<J_1,J_2\ll |J_0|$, the limit of weakly
coupled Ising chains.
Our focus is on the effects of quantum fluctuations controlled by the
transverse field and hence we will not address early experimental
work~\cite{kobayashi,heid} which studied the material in a longitudinal
field. Our results with some modifications also apply to other materials
which are known to realize Ising spin chains weakly coupled on a
triangular lattice~\cite{cao}. Theoretically, a study of the
2-dimensional triangular Ising anti-ferromagnet in a transverse field is
available~\cite{ms}, which corresponds to the limit of $J_{0}=0$ of our
model. We are interested in the opposite limit in this paper.  Were
$\alpha\equiv J_2/J_1=1$, the system would form a perfect triangular
lattice in the basal plane. It is known that the material has $\alpha
\lesssim 1$ and hence the coupling along the chains running along the
x-axis is a little stronger~\cite{kobayashi}. As we shall see, the
lattice anisotropy significantly enriches the phase diagram.

When $J_1,J_2=0$ the model Eq.~(\ref{eq:model}) reduces to a
collection of one-dimensional Ising chains in a transverse field
(TFIC),
\begin{equation}
\label{eq:tfic}
H_{\rm TFIC} = J_0 \sum_{i} S^z_i S^z_{i+1}-h_\perp\sum_i
S^x_i -h_\parallel\sum_iS^z_i.
\end{equation}
An $h_\parallel$ term is included even though we study the model
Eq.~(\ref{eq:model}) with no parallel field; the reason for this will become clear
below. The TFIC harbors a quantum critical point (between a magnetic and
a paramagnetic state when $h_\parallel=0$ and $h_\perp=J_0/2$) in the same
universality class as the well studied two-dimensional Ising field
theory. 
Indeed, the singular contribution to the ground state energy of the TFIC
Hamiltonian, Eq.~(\ref{eq:tfic}) can be related to the free
energy of the Ising field theory by a simple re-scaling
of variables. The scale is set by comparing correlation functions in
the TFIC (it can be solved exactly when $h_\parallel=0$, see {\em e.g.}~\cite{pfeuty}) to the Ising
field theory results~\cite{fz}. We find,
\begin{equation}
E_{\rm TFIC}= \frac{J_0}{2} \mathcal{E}_{\rm IFT}\left (\frac{
  h_\perp-J_0/2}{J_0/2},
\frac{c_h h_\parallel}{J_0/2 }\right ).
\end{equation}
We will make use of the scaled variables $\bar h_\perp =
\frac{h_\perp-J_0/2}{J_0/2} $, $\bar h_\parallel=\frac{c_h h_\parallel}{J_0/2 }$ and express all energies in
units of $J_0/2$ ($c_h = \sqrt{\frac{e^{1/4} 2^{1/12}}{4
    A^3}}\approx  0.4016$, $A\approx 1.2824$ is Glaisher's
constant). Expansions for the function ${\cal E}_{\rm IFT}(\bar
h_\perp,\bar h_\parallel)$ are available~\cite{fz} in
different limits for both finite $\bar h_\perp$ and $\bar h_\parallel$,  when
$\xi=\bar h_\parallel/\bar h_\perp^a \ll 1$  and when the magnitude of
$\eta=-\bar h_\perp/\bar h^{1/a}_\parallel$ is small,
\begin{eqnarray}
\label{eq:fz}
{\cal E}^{\bar h_\perp>0}_{\rm IFT}(\bar h_\perp,\bar h_\parallel)&=&
\frac{\bar h_\perp^2}{8\pi} \log \bar h_\perp^2+ \bar h_\perp^2\left (
G_2\xi^2+G_4\xi^4+G_6\xi^6\dots \right),\nonumber\\
{\cal E}^{\bar h_\perp<0}_{\rm IFT}(\bar h_\perp,\bar h_\parallel)&=&
\frac{\bar h_\perp^2}{8\pi} \log \bar h_\perp^2+ \bar h_\perp^2\left (
\tilde G_1 |\xi| + \tilde G_2 |\xi|^2+\dots \right ),\nonumber\\
{\cal E}_{\rm IFT}(\bar h_\perp,\bar h_\parallel)&=&
\frac{\bar h_\perp^2}{8\pi} \log \bar h_\perp^2+ \bar h_\parallel^{16/15}\left[
  -\frac{\eta^2}{8\pi}\log \eta^2+\left(\Phi_0+\Phi_1 \eta + \Phi_2\eta^2+\dots \right)\right],
\end{eqnarray}
where
$a=15/8$ is a critical index of the Ising universality class. The
numbers $G_n$, $\tilde G_n$ and $\Phi_n$ are universal numbers which are known
to great accuracy, a few are reproduced in Table~\ref{table:Gn}.

In order to expand around the non-trivial TFIC limit we construct a
variational wave-function, $|\Psi \rangle = \prod_i |\bar h_i\rangle$,
as a product of TFIC ground states, where the longitudinal field $\bar
h_\parallel = \bar h_i$ defining the TFIC Hamiltonian for each chain is
taken as a variational parameters. The variational energy for our model
is given by,
\begin{eqnarray}
E_{\rm mf}&=&\sum_{i} \mathcal{E}_{\rm IFT}(\bar h_\perp, \bar h_i)
+\bar h_i\bar m_i + \sum \bar m_i
\frac{\bar J_{ij}}{2} \bar m_j,
\label{eq:var_free}
\end{eqnarray}
where the magnetization on the $i^{\rm th}$ chain, $\bar m_i=-\partial
\mathcal{E}_{\rm IFT}/\partial \bar h_i$. Optimization over $\bar h_i$, gives
$\bar h_i=- \bar J_{ij}\bar m_j$
\begin{eqnarray}
E_{\rm mf}&=&\sum_{i} \mathcal{E}_{\rm IFT}(\bar h_\perp, \bar h_i) - \sum \bar h_i
\frac{\bar J_{ij}^{-1}}{2} \bar h_j.
\label{eq:mf_free}
\end{eqnarray}
Here the labels $i,j$ span sites in a plane (or corresponding chains),
and $J_{ij}$ is the interaction between a pair of spins.  In addition we
have defined the rescaled matrix $\bar J_{ij}\equiv c_h^2
J_{ij}/(J_0/2)$, whose non-zero entries are $\bar J_{1,2}=c_h^2
J_{1,2}/(J_0/2)$.  Sometimes we prefer to work with the mean-field
magnetization, $\bar m_i=\bar J^{-1}_{ij}\bar h_j$, in which case
$E_{\rm mf}=\sum_{i} \mathcal{E}_{\rm IFT}(\bar h_\perp,- \bar
J_{ij}\bar m_j) - \sum \bar m_i\frac{\bar J_{ij}}{2} \bar m_j$. The
philosophy of our approach is similar to ``chain mean-field'' theories
that has been applied to a number of unfrustrated
problems~\cite{schulz,tsvelik,scalapino}.

We now study the quantum phase diagram (phases and phase transitions) that arise
by extremizing the variational energy Eq.~(\ref{eq:var_free}) using the exact
expansions Eq.~(\ref{eq:fz}) for the TFIC. We present our results for the
phase diagram as a functions of $\bar J_1$ and $\bar h_\perp$, for a fixed $\alpha$.

{\em Perfect Triangles:}
When $\bar J_2=\bar J_1$ ({\em i.e.}~$\alpha=1$), we have perfect
equilateral triangles in
the basal plane. Treating this coupling in perturbation theory, we
approach the magnetic ordering from each of the two phases of the
Ising chain, the paramagnetic phase for $\bar h_\perp>0$ and the magnetic
phase for $\bar h_\perp<0$.

\begin{figure}[h]

\subfigure[Ferri-magnetic ({\bf FR})]{
      \includegraphics[width=1in,angle=-90]{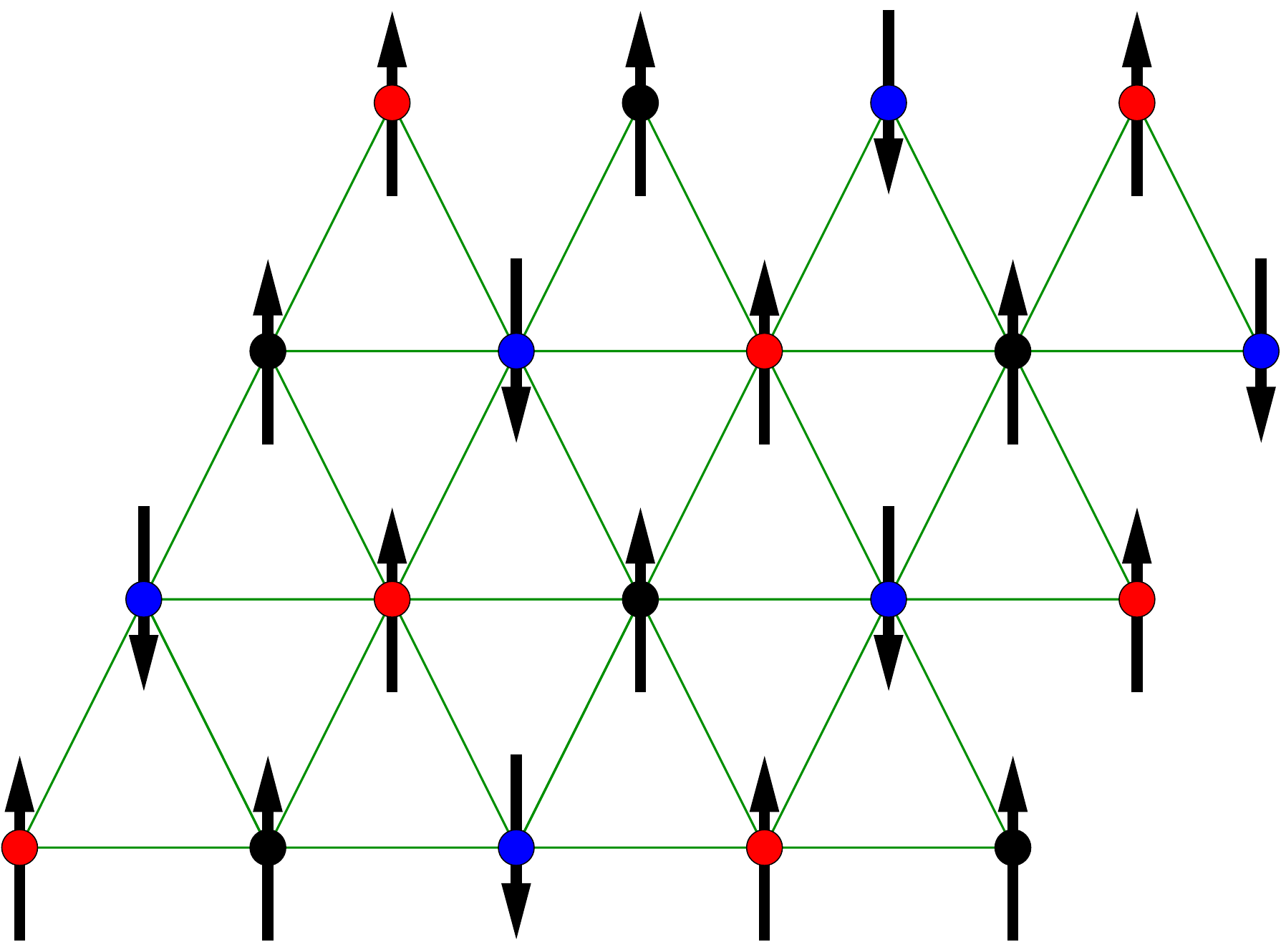}
    }

\subfigure[Anti-Ferromagnet ({\bf AF})]{
      \includegraphics[width=1in,angle=-90]{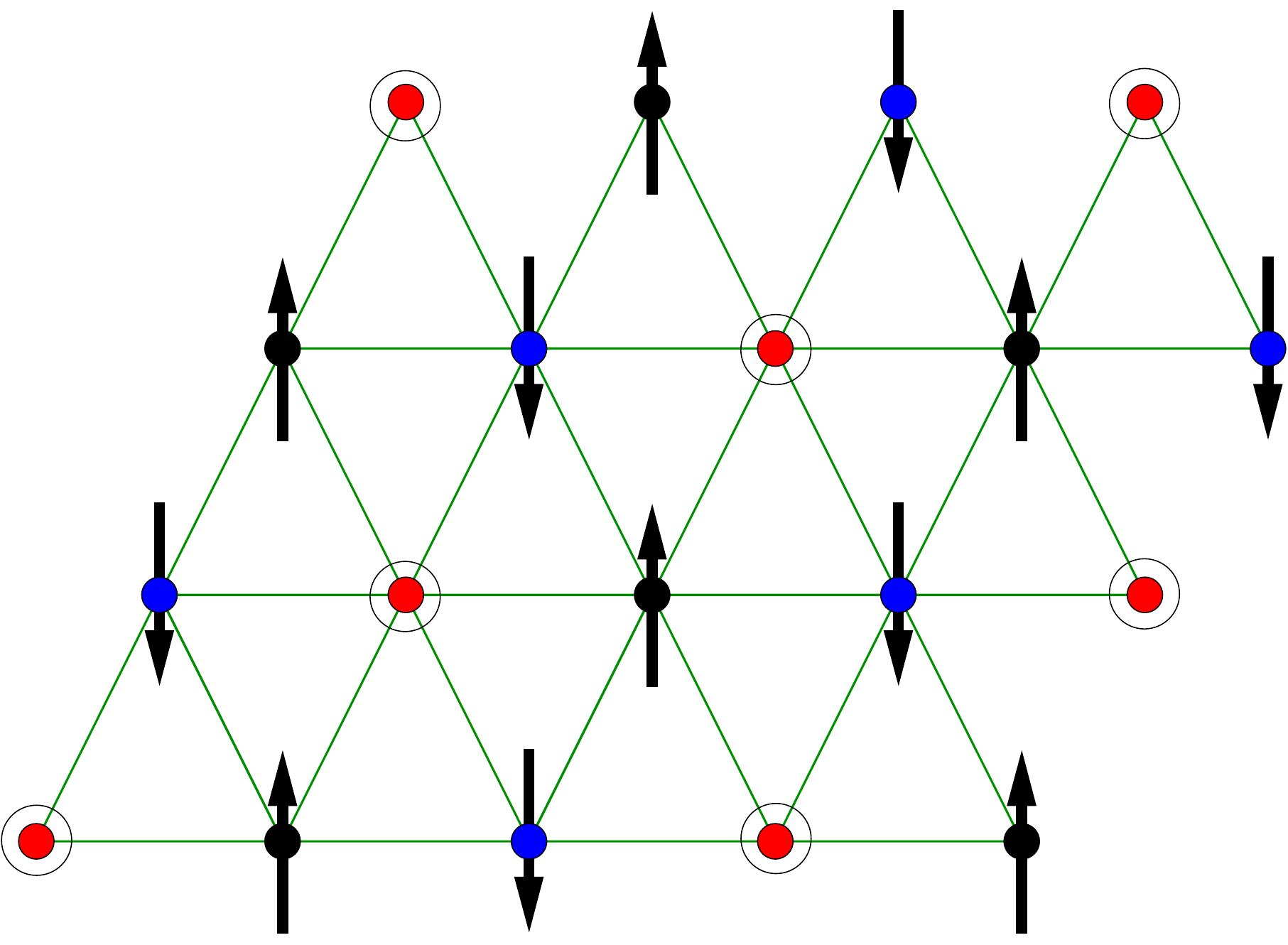}
    }

\subfigure[N\'eel 1 ({\bf N1})]{
      \includegraphics[width=1in,angle=-90]{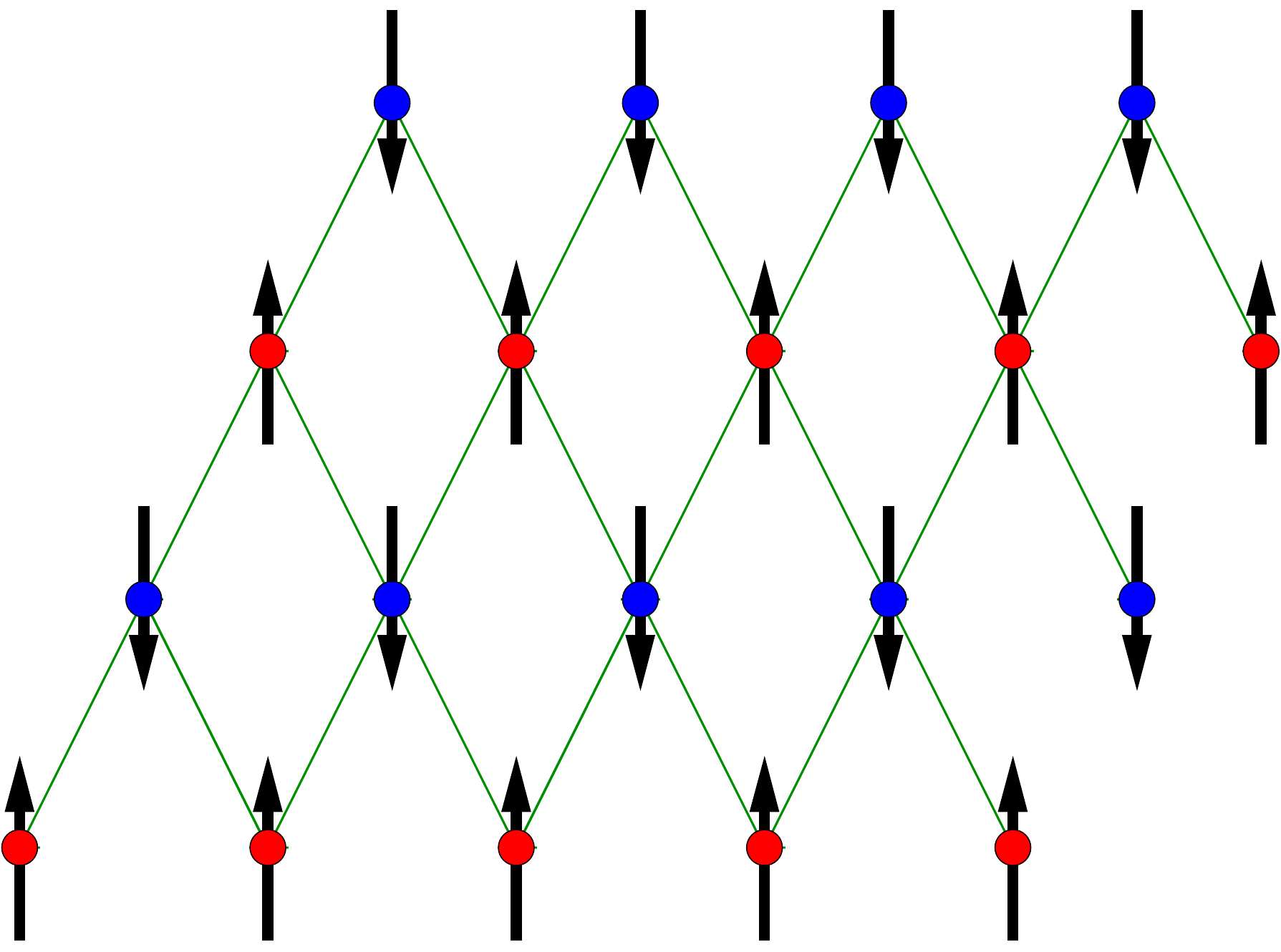}
    }

\subfigure[N\'eel 2 ({\bf N2})]{
      \includegraphics[width=1in,angle=-90]{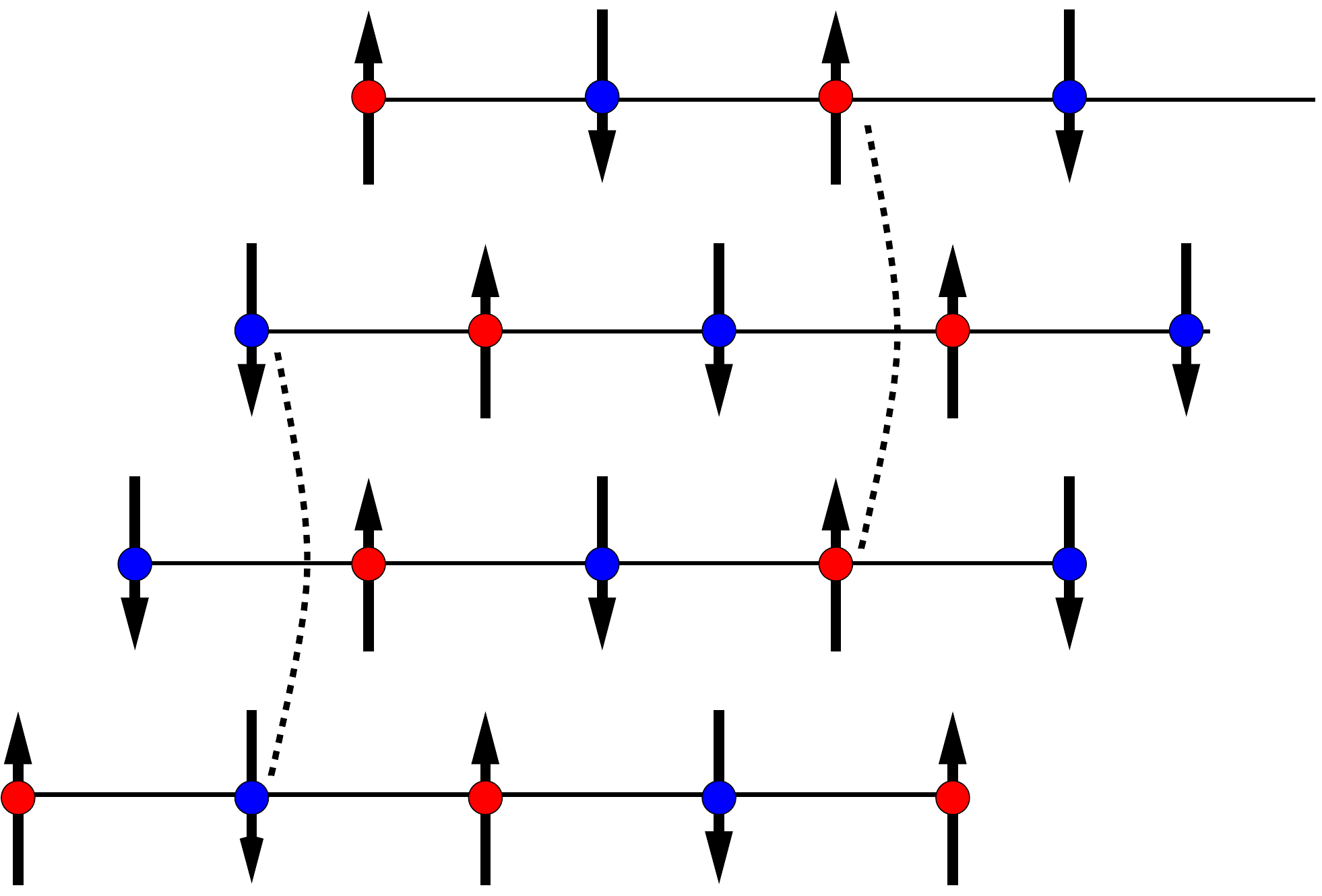}
    }
    \caption{\label{fig:mag_order}Magnetic ordering patterns of {\bf FR,
        AF, N1 and N2} phases. (a) Dividing the lattice into A, B and C
      sub-lattices (marked red, blue and black), the {\bf FR} phase
      corresponds to putting the spins on two of these parallel and the
      spins on the third anti-parallel, leading to a six-fold
      degeneracy. (b) The {\bf AF} state is also a three sub-lattice
      state but with the spin $\langle S_i^z\rangle$ on one site
      fluctuating to zero (c) the two-fold degenerate {\bf N1} phase,
      obtained when $\alpha>1$, can be thought of most simply as the
      bi-partite (red and blue) N\'eel phase obtained when the $J_1$
      bonds are set to zero. (d) the four-fold degenerate {\bf N2} phase
      obtained when $\alpha<1$ corresponds to a N\'eel ordering along
      the basal chains ($J_1$ bonds) and a ferromagnetic locking of
      alternate chains (shown by dashed lines).  }
\end{figure}

When $\bar h_\perp < 0$, each isolated chain has a two-fold degenerate
magnetic state. Once the chains are coupled this degeneracy is lifted by
the exchange fields from the neighbors. Working perturbatively, we
expand the energy Eq.~(\ref{eq:mf_free}) order by order in $\bar J_1$
till the degeneracy is fully lifted.  At leading order after minimizing
the energy with respect to the magnitude of $\bar m_i$, $E_{\rm
  mf}\approx \bar m_i\frac{\bar J_{ij}}{2} \bar m_j$, (with $|\bar m_i|$
fixed) i.e. we obtain the classical triangluar lattice Ising model for
the signs of $\bar m_i$. It is well known that there is an extensive
degeneracy of configurations which minimize the energy\cite{Wannier},
i.e. ground states consist of all configurations with exactly one pair
of parallel spins ($\bar m_i$ of the same sign) on each elementary
triangle.  We must continue expanding in $\bar J_1$ to find how this
entropy is finally lifted. At next order, the free energy receives a
contribution, $\delta E_{\rm mf} \approx \tilde G_2 \sum_i |\bar
h_\perp|^{2-2a}( \bar J_{ij}m_j)^2$.  Since $\tilde G_2<0$, this
perturbation prefers to have spins on all elementary hexagons
parallel. So finding the quantum ground state of our model when $\bar
h_\perp<0$ is reduced to finding the classical ground state of the
model, $H = J\sum_{\langle ij \rangle}\sigma_i \sigma_j - K\sum_i \left
  ( \sum_{a}\sigma_{i+e_a}\right )^2$ where $e_a$ are all the neighbors
of a site, when $K, J>0, K\ll J$ and $\sigma_i=\pm1$. We find that the
ground state (proven in ~\ref{sec:JK_model}) is a three-sublattice
ferrimagnetic state with spins pointing up on the $A$ sub-lattice and
the spins pointing down on the B and C sub-lattices (there are a total
of 6 symmetry-related states). We denote this state as {\bf FR}, see
Fig.~\ref{fig:mag_order}.

When $\bar h_\perp >0$, the isolated chains are in a
non-degenerate gapped paramagnetic ({\bf PM}) state. Perturbation theory in the
coupling $\bar J_1$ leaves this state stable and hence there is always a
region of {\bf PM} for small enough $\bar J_1$. We can now ask if we increase
$\bar J_1$ what magnetic order will first condense. 
The primary condensation takes place at
$(4\pi/3,0)$.   Since the order parameters are small, in this regime we
can perform a Landau expansion for the order parameter. We have to
consider both the complex order parameter at ${\bf Q_c}=(4\pi/3,0)$ which we
call $\Phi$ and the real uniform component at $(0,0)$ which we call
$m$: $h_i=\Phi e^{\bf iQ_c\cdot r}+\Phi^* e^{\bf- iQ_c\cdot r} +
m$. Keeping only the terms relevant to our study from the most general
expansion for these two fields, we arrive at the following Landau continuum
expression for the energy density $E_{\rm mf}\approx\int d^2 r\epsilon_{\rm L}$, ignoring gradient terms,  
\begin{equation}
\label{eq:free_lambda}
\epsilon_{\rm L} = \alpha_\Phi|\Phi|^2+\beta_\Phi|\Phi|^4 + \alpha_mm^2 + \lambda_3 m (\Phi^3 +\Phi^{*3})+\lambda_6(\Phi^6+\Phi^{*6})
\end{equation}
The coeffecients are fixed by the parameters,
$G_2,G_4,G_6,\bar h_\perp$ and $\bar J$. As $\bar J_1$ is increased, the condensation of $\Phi$ takes
place when $\alpha_\Phi=0$ (note however $\alpha_m$ remains non-zero) in
Eq.~(\ref{eq:free_lambda}),
when $\bar J^c_1=\frac{1}{6|G_2|}\bar h_\perp^{2a-2}$.
We
can then optimize $m$  for a given configuration
of $\Phi$, $m=-\lambda_3/(2\alpha_m)(\Phi^3+\Phi^{*3})$ and eliminate
it from the energy.
This shifts the $\lambda_6$ term to $\lambda_6-\lambda_3^2/(4\alpha_m)$,
which evaluates to $\frac{3G_2G_6-8G_4^2}{3G_2}\bar h_\perp^{2-6a} > 0$
at the critical point.  This favors states in which $\Phi^6$ is real and
negative, so that the phase $\theta$ defined by $\Phi=|\Phi|e^{i\theta}$
is pinned to $\pi/2$. Magnetic order with $\Phi=|\Phi|e^{i\pi/2}$ and
$m=0$ corresponds to an anti-ferromagnetic state, we call ${\bf AF}$,
illustrated in Fig.~\ref{fig:mag_order} (b). This establishes that the
${\bf AF}$ order forms first when the critical $\bar J^c_1$ is crossed.

The pattern of magnetic ordering close to $\bar h_\perp=0$ can be found using the
$\eta$-expansion. In this regime, we find with $\eta=0$ that
the {\bf FR} state has the lowest energy and this state can be smoothly connected to
the state found for $\bar h_\perp<0$.

\begin{figure}
  \includegraphics[width=8in]{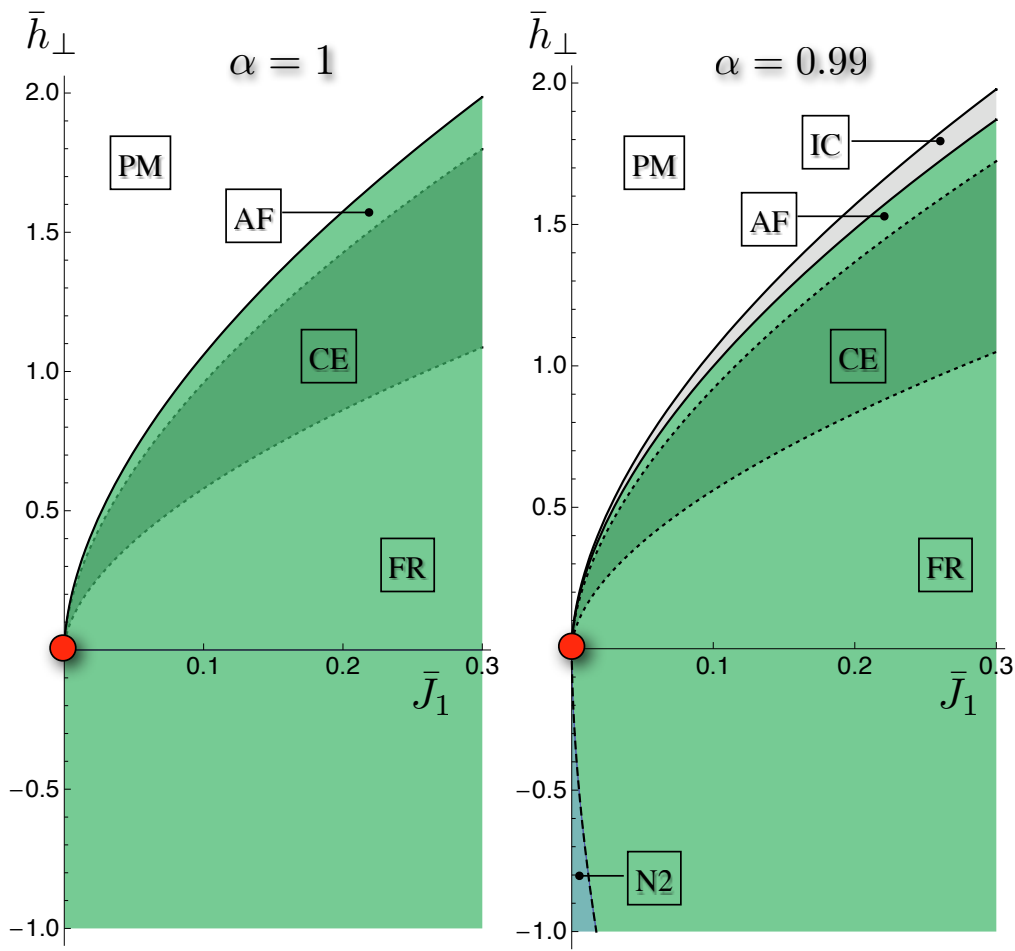}
  \caption{ \label{fig:pd}Calculated phase diagram for model, Eq.~(\ref{eq:model}) for
    perfect triangles, $\alpha\equiv J_2/J_1=1$, (left) and with an isosceles distortion
    $\alpha=0.99$ (right). The red dot indicates the Ising quantum
    critical point of the decoupled chains. The region marked {\bf CE}
    is a co-existense between the {\bf AF} and {\bf FR states}. It was
    estimated by using a numerical interpolation for the different
    expansions of the scaling form of the energy.  }
 
\end{figure}

Putting together the results from the three expansions, when $\bar
h_\perp>0$, $\bar h_\perp<0$ and $\bar h_\perp=0$, we find that the
simplest form of the phase diagram consistent with these expansions is
the one shown in Fig.~\ref{fig:pd}(a).   On general grounds, the
transition between ${\bf FR}$ and ${\bf AF}$ can either be direct and
and first-order, or it may proceed through a region of ``coexistence''
corresponding to states with complex $\Phi^6$. 

{\em Isosceles Triangles:} Now we adapt and extend the above analysis
to the more complicated case when $\alpha= \bar J_2/\bar J_1 \neq 1$, resulting
from an isosceles distortion of the
triangular lattice in the basal plane 
(as occurs in the real material). 

When $\bar h_\perp<0$, once again there is a macroscopic degeneracy
associated with the spontaneous symmetry breaking on each chain which
will be lifted by the $\bar J_{1,2}$ interactions. Exactly as in the
$\alpha=1$ case, at leading order, we have an Ising model for $m_i$, but
now on an anisotropic lattice, $E_{\rm mf}= \bar m_i\frac{\bar
  J_{ij}}{2} \bar m_j$. For $\alpha>1$ the ground state is a two-fold
degenerate N\'eel state ({\bf N1}), shown in
Fig.~\ref{fig:mag_order}. For $\alpha<1$ (the case in CoNb$_2$O$_6$),
the situation is more complicated, and at leading order the ground state
still has a degeneracy. The triangular lattice in the basal plane can be
divided into ``basal'' chains (not to be confused with the c-axis Ising
chains), see Fig.~\ref{fig:isosceles}, each of which is N\'eel ordered,
however each basal chain can be shifted by a unit without lifting the
degeneracy.  In fact, in mean field theory this sub-extensive degeneracy
is not lifted at all, an artifact of this approximation.  Instead, a
full perturbative expansion splits the degeneracy at fourth order (see
Appendix~\ref{app:pert}), by creating a ferro-magnetic interaction
between alternate basal N\'eel chains (even and odd basal chains get
locked separately).  This effective interaction is a quantum fluctuation
effect, not present in the classical $h_\perp=0$ limit.  The phase shift
between even and odd basal chains is determined spontaneously and leads
to an extra 2-fold degeneracy, resulting in a 4-fold degeneracy for the
{\bf N2} state.  For a fixed $\bar h_\perp<0$ and $1-\alpha \ll 1$, for
arbitrarily small $\bar J_1$, we have shown that the {\bf N2} state is
selected.  If $\bar J_1$ is made large enough however one expects a
transition back into the {\bf FR} state.  The phase boundary between
{\bf N2} and {\bf FR} is first order, and can be estimated by asking how
big $\bar J_1$ should be so that the $\delta E_{\rm mf} \approx \tilde G_2 \sum_i |\bar
h_\perp|^{2-2a}( \bar J_{ij}m_j)^2$ (derived above for $\alpha=1$) term that picks the {\bf FR} state
 is large enough to close the gap that the anisotropy opens; we
find $\bar J_1^{\rm N2-FR}=\frac{\bar h_\perp^{2a-2}(1-\alpha)}{2|\tilde
  G_2| ((1+2\alpha)^2-3)}$.

When $\bar h_\perp>0$, we proceed as in the case of perfect
triangles. For small $\bar J_1$ the {\bf PM} state is stable.  However,
at sufficiently strong coupling there is again a condensate but it
occurs at an {\em incommensurate} wave-vector. The onset wave-vector is
determined completely by $\alpha$, it is $(q_x^*,0)$, where
$q_x^*=2\cos^{-1}\left(-\frac{\alpha}{2}\right)$, shifted from
$(4\pi/3,0)$. The phase transition will occur again when the coefficient
of the quadratic term of the $(q_x^*,0)$ boson mode changes sign, this
determines the critical coupling: $\bar J^c_1=\frac{\bar
  h_\perp^{2a-2}}{2|G_2|(2+\alpha^2)}$. We call this phase with
incommensurate magnetic ordering ${\bf IC}$.  As $\bar J_1$ is made even
stronger the system will eventually prefer to lock-in to a commensurate
ordering vector. For $1-\alpha \ll 1$, this state will be the {\bf AF}
state discussed above.  To understand the commensurate-incommensurate
phase transition, it is convenient to begin in the commensurate ${\bf
  Q_c}=(4\pi/3,0)$ {\bf AF} phase. We can then understand the transition
to the incommensurate state as the formation of a dilute condensate of
domain walls (which separate the 6 degenerate minima). We begin by
evaluating the mean-field energy, for configurations of the form $\bar
h_i = |\Phi|\left [ e^{i({\bf Q_c\cdot r_i} + \theta({\bf r_i}))}+ {\rm
    c.c.}\right ] + m$, where $\theta({\bf r})$ is a smooth field which
creates a domain wall when it shifts between domains, $\theta
=(2\pi/6)n$.   Owing to the anisotropy of the problem, we may assume
that (apart from fluctuations), $\theta({\bf r})$ depends only upon
$x$.  After coarse-graining and extremizing with respect to $m$, we then
obtain the famous sine-Gordon
model~\cite{chaikin},
\begin{equation}
\label{eq:sg_model}
E_{\rm sg}= A_{yz}\int dx \left [\frac{\kappa}{2}(\partial_x \theta)^2  +
  \delta_x \partial_x \theta + \lambda \cos(n_{\rm sg}\theta)\right ],
\end{equation}
where $A_{yz}$ is the sample area in the $y-z$ plane, and the couplings
of the SG model can be related to the parameters in
Eq.~(\ref{eq:mf_free}) as worked out explicitly in
Appendix~\ref{app:cit}; $n_{\rm sg}=6$.  
From these expressions,
we can calculate the phase boundary between the commensurate {\bf AF} state
and the {\bf IC} state at leading order in $(1-\alpha)$: 
$\bar J^{\rm CIT}_1 = \bar J^c_1+\mathcal{A}_{\rm CIT} \bar h_\perp^{2a-2} (1-\alpha)$,
where $\mathcal{A}_{\rm CIT}=\frac{G_4 \pi}{-2G_2\sqrt{16G_4^2-6G_2G_6}}$ is a
pure number.

Our arguments above have predicted a phase diagram shown in Fig.~\ref{fig:pd}
when $1-\alpha \ll 1$. As expected from scaling arguments, all the phase boundaries scale as
$\bar J \approx \mathcal{A} h^{2a-2}_\perp$ when the TFIC's are close to their individual
critical points, where $\mathcal{A}$ is an amplitude which depends on
the transition being considered. We have succeeded in computing the
amplitudes, $\mathcal{A}$ in some rather non-trivial approximations which we have detailed
in the body of the work. Since $\alpha\lesssim 1$ in CoNb$_2$O$_6$, our
results should apply to this material.  
Let us now turn to a discussion of the experimental response of
each the phases as $\bar h_\perp$ is increased from zero and to the nature of the phase
transitions between them.

{\em N\'eel phase:} This is the phase {\bf N2} (illustrated in
Fig.~\ref{fig:mag_order}) which appears with weak external fields. It
has no net magnetization and exhibits an elastic neutron peak at
momentum $(\tfrac{\pi}{2},0)$ ($(0,\tfrac{1}{2},0)$ in the orthorhombic
co-ordinate system).  This is in agreement with zero field
experiments\cite{kobayashi,heid}, though further-neighbor interactions
not present in our model could play a role at $h_\perp=0$.

{\em Ferri-magnetic phase:} As $\bar h_\perp$ is increased, we expect a
first order phase transition into the {\bf FR} state. The transition
{\bf N2-FR} is driven by the increase of quantum fluctuations of the
Ising spins. The {\bf FR} phase has a net uniform magnetic moment and
thus is characterized by elastic neutron peaks both at zero momentum and
at the commensurate ordering vector $(4\pi/3,0)$.  As $\bar h_\perp$ is
increased we expect the {\bf FR} order to weaken causing a reduction of
the Bragg peaks; the uniform magnetization $m$ and hence zero momentum
peak goes to zero as $\bar h_\perp$ is increased much faster (like the
third power) than the peak at $(4\pi/3,0)$.  

{\em Anti-Ferromagnetic phase:} The {\bf AF} state also orders at
$(4\pi/3,0)$, but it has no uniform component of the magnetization in
the z-direction and is hence analogous to an anti-ferromagnet. It is
entirely different from the {\bf N2} state however, and is truly a
result of the geometric frustration present.   The difference between
the ${\bf AF}$ and ${\bf FR}$ states could be probed by tilting the
field in the $b-c$ plane -- one expects a weak first order transition at
zero tilt in the ${\bf FR}$ case but not the ${\bf AF}$ one.

{\em Incommensurate:} As $\bar h_\perp$ is increased further still,
there is a transition into the {\bf IC} phase that is driven by a
condensation of domain walls in the {\bf AF} phase.  The {\bf IC} phase
is characterized by a dense set of Bragg peaks and a massless goldstone
(phason) mode.  The primary ordering vector, $q_x$, shifts continuously in
this phase from  $q_x=4\pi/3$ in the {\bf AF} phase to
value $q_x^*=2{\rm cos}^{-1}(-\alpha/2)$ at the transition to the {\bf
  PM},   the evolution of $q_x(\bar h_\perp)$ is described in
~\ref{sec:qstar_shift}).   

{\em Paramagnetic:} Finally when the quantum fluctuations due to
$\bar h_\perp$ are strong enough, magnetic ordering is completely suppressed
and the system enters a quantum paramagnetic phase. This state is
gapped and breaks no symmetries of the model, Eq.~(\ref{eq:model}).

{\em Quantum Phase Transitions:} There are four quantum phase
transitions that we have predicted in the evolution of phases as $\bar
h_\perp$ is tuned. Although the location of these transitions relies on
mean field theory, their critical properties are universal.  The first
transition between {\bf N2-FR} is expected to be strongly
first-order. The second transition is between the {\bf FR-AF} states. It
can take place either by a direct first order transition or through a
coexistence of the two phases.  Within mean field theory, we find the
latter, though this conclusion may be sensitive to the approximation,
and to small perturbations of the Hamiltonian.   The third transition
between the {\bf AF-IC} state is continuous and is of the
commensurate-incommensurate type. We have calculated the leading
singularity of the ordering vector here.   The final and fourth transition
is between {\bf IC-PM}, which we expect to be in the XY universality
class in 4 classical dimensions. Since this is at the upper critical
dimension we expect mean field exponents with log corrections. 
The transverse magnetization as a function of the external field,
$h_\perp$, should provide a good method for locating the phase transitions.
The magnetization itself should have discontinuities at the first order transitions and
it's first derivative should have singularities at the locations of
the second order transition.

In summary, we have studied the model, Eq.~(\ref{eq:model}), for
CoNb$_2$O$_6$ in an external magnetic field that is aligned along the
crystallographic b-axis, which results in a fully quantum mechanical
``transverse field''. Application of the transverse field drives the
frustrated Ising model under study into a number of fascinating phases
through quantum phase transitions. We have outlined the basic properties
of each of these phases here.  Detailed calculations of the excitation
spectrum, measurable by neutron scattering, are also possible, and will
be presented in a follow-up publication.  An interesting question we
leave for future exploration is the nature of the finite-temperature
phase diagram of this model.

{\em Acknowledgements.-} We thank Radu Coldea for stimulating
discussions and acknowledge support from the
Packard Foundation and National Science Foundation through grants
DMR-0804564 and PHY05-51164.

\section{Supplementary Materials}

\subsection{Ground state for $\alpha=1$ and $\bar h_\perp<0$}
\label{sec:JK_model}

Here we prove that the ground state of the model on the perfect
triangular lattice, $H = \bar J\sum_{\langle
  ij \rangle}\sigma_i \sigma_j - K\sum_i \left (
  \sum_{a}\sigma_{i+e_a}\right )^2$, for $\bar J,K>0$ and $K/\bar J \ll 1$ is
the {\bf FR} state.
The proof is as follows: We need to find which
Ising configuration has the lowest energy with the $K$ term. Lets
imagine an Ising configuration on an infinite lattice with
$f_6,f_5,f_4,f_3$ the fraction of sites that have $6,5,4,3$
anti-parallel neighbors. Then $f_6+f_5+f_4+f_3=1$. Also a third of the
bonds have to be frustrated (host parallel spins) in any Ising
configuration which implies, $f_5+2f_4+3f_3=2$. The contribution of
the $K$ term to the energy in this language is: $E= - K (36 f_6+16 f_5
+4f_4)$. Since we have two extra equations, and four unknowns, we can
rewrite the equation for $E$ completely in terms of $f_3$ and $f_6$,
obtaining $E= -K (4+ 8(f_3+f_6))$. Since $f_3+f_6<1$, the lowest
energy is clearly obtained only with $3,6$ sites. This configuration
has $f_6=1/3$ and $f_3=2/3$. The only such configuration is the
ferri-magnetic state described above.

\subsection{Ground state for $\alpha<1$ and $\bar h_\perp<0$}
\label{app:pert}

In the section we show that the ground state when
$\alpha<1$ and when the chains are magnetically ordered is the {\bf N2}
state for arbitrarily small inter-chain coupling.
The goal here is to develop a systematic perturbative expansion about the limit of
decoupled Ising chains. On the magnetic side there is a huge
degeneracy in this limit because each chain can have its magnetization $m_i$ point
up or down. Working in imaginary time, we can set it up as follows,
\begin{equation}
\frac{Z}{Z_0}=\frac{\int e^{-S_{\rm 1d}-S_{\rm c}}}{\int e^{-S_{\rm
      1d}}}=\langle e^{-S_{\rm c}} \rangle,
\end{equation}
where $S_c=\int d\tau S^z_i(\tau) \frac{\bar J_{ij}}{2}S^z_j(\tau)$ and all
averages are taken over the decoupled limit. Note that in the zero
temperature limit, the partition function is simply given by, $Z=e^{-\beta E_{\rm gs}}$. We can calculate the average perturbatively using the cumulant expansion,
\begin{equation}
\langle e^{-S_{\rm c}} \rangle = e^{-\left[  \langle S_{\rm c}\rangle
    - \frac{1}{2!}\langle S^2_{\rm c}\rangle_{\rm con} 
    + \frac{1}{3!}\langle S^3_{\rm c}\rangle_{\rm con} 
    - \frac{1}{4!}\langle S^4_{\rm c}\rangle_{\rm con} 
    +\dots   \right]},
\end{equation}
where the subscript ``con'' indicates that we should only keep
connected averages. When we have to evaluate the connected averages we
can factor the average into products of averages each of which acts
within a single chain, since the chains are decoupled at $0^{\rm th}$
order. For an odd number of spins, these single chain averages will be
replaced by, e.g. $\langle S^z_i(\tau)\rangle \rightarrow \bar m_i |\langle
S^z(\tau) \rangle|$ with $\bar m_i=\pm 1$. Then we can ask which
configuration of $\bar m_i$ produces the lowest energy and keep evaluating
higher order contributions till any ``accidental'' degeneracy is completely
lifted. The coeffecients of the interactions between $\bar m_i$ are determined by integrals over
correlation functions of the d=1 Ising chain.

We now turn to an evaluation of this approach for the case of
interest, when $\alpha <1$. {\em At leading order}, we have the term $\bar m_i
\frac{\bar J_{ij}}{2} \bar m_j$, which gives N\'eel configurations along the
stronger bonds that form a
chain in the basal plane (not to be confused with the Ising chains in
the transverse direction!), but each basal chain can be translated by a unit leading
to a $2^{N_{\rm chain}}$ degeneracy (sub-extensive entropy). We expect
that the degeneracy is lifted by locking-in of spins on alternate basal
chains, but we still need to determine whether the locking-in is
ferromagentic or anti-ferromagnetic. {\em At second order}, we generate
ferro two-spin interactions between alternate basal chains $\propto - \bar J_{ij}\bar J_{ik}
\bar m_j m^k$ but it is easy to see
that these cancel when summed over the sites of the basal chain due to
the N\'eel order,
{\em i.e.,} the mean field from all the sites in a chain add up to
zero. The degeneracy is not lifted at this order. {\em At third
order}, we generate both two-spin and four-spin couplings. It turns
out that both the two- and four-spin couplings cancel out. {\em At
  fourth order}, we have a large number of generated two-spin couplings and
four-spin couplings. Turns out that the four-spin coupling does not
lift the degeneracy, but a careful summation of the 86 two-spin couplings does!
The two-spin interactions at fourth order are ferromagnetic and
hence alternate chains are locked ferromagnetically as shown
in Fig.~\ref{fig:mag_order}.

\subsection{Commensurate-Incommensurate Transition}
\label{app:cit}

In this section we calculate the energy of a domain wall in the
commensurate order parameter as a function of the external parameters. The simplest
commensurate-incommensurate transition (CIT) occurs when single
domain walls condense~\cite{chaikin}, i.e. when the energy of a single
domain wall goes to zero. We can estimate this by a calculation of the domain
wall energy. A domain wall corresponds to a spatial profile of the
phase $\theta({\bf r})$ of the complex order parameter $\Phi$
introduced earlier in the context of Eq.~(\ref{eq:free_lambda}).  We hence consider the following form for the field $\bar h_i$ at site $\bf{r}_i$,
\begin{equation}
 \bar h_i= |\Phi|\left [e^{i({\bf Q_c} \cdot {\bf r}_i + \theta({\bf
       r}_i))}+{\rm c.c.}\right ] +m.
\label{eq:m_state}
\end{equation}
Here, ${\bf Q}_c$ is a commensurate wave vector $(4 \pi/3,0)$ and
$\theta({\bf r})$ is a slowly varying functions of
{\bf r}. We substitute this expression into our variational energy,
Eq.~(\ref{eq:var_free}) and then derive a coarse grained energy density for the
field $\theta({\bf r})$. Next we integrate out $m$, like we discussed
below Eq.~(\ref{eq:free_lambda}), arriving at the Sine-Gordon
long-wavelength energy for $\theta$, Eq.~(\ref{eq:sg_model}), where we
have:
\begin{eqnarray}
\label{eq:cg_cplg}
\lambda_6&=& \frac{-8G^2_4+G_2G_6}{G_2}\bar h_\perp^{2-6a},\nonumber \\
\lambda_3&=& -8G_4\bar h_\perp^{2-4a},\nonumber \\
\alpha_{\rm m} &=& -3 G_2 \bar h_\perp^{2-2a}, \nonumber \\
\lambda&=&2 |\Phi|^6 \left ( \lambda_6 - \frac{\lambda_3^2}{4\alpha_{\rm m}}\right
),\nonumber \\
\kappa_x&=& - |\Phi|^2G_2 \bar h_\perp^{2-4a},\nonumber \\
\delta_x&=& -|\Phi|^2\frac{2G_2(1-\alpha)}{\sqrt{3}}h_\perp^{2-2a}.\nonumber
\end{eqnarray}
Except for $\delta_x$, we evaluated all these couplings with $\alpha=1$.

Now we are all set to calculate the transition from the state with commensurate
$(4\pi/3,0)$ ordering to the incommensurate phase. Since the $\delta$
term has only an $x$ gradient, we will look for
solutions that are translationally invariant in the $y-$direction, dropping all derivatives in this direction. Noting that
$\delta$ only gives a boundary term, it may expressed simply as the
winding number $N_w=[\theta(x=L)-\theta(x=0)] \frac{1}{2\pi/6}$, the
number of times the field $\theta$ moves from one its six minima to
an adjacent one. The solution for $N_w=\pm 1$ is well known,
\begin{equation}
\theta(x)=\frac{4}{n} \tan^{-1} (e^{\pm n_{\rm sg} \sqrt{\frac{\lambda}{\kappa_x}} (x-x_0)}),
\end{equation}
where $n_{\rm sg}=6$ in our case.

The energy of a single domain wall is positive for $|\delta|<|\delta_c|=4 \sqrt{\kappa_x \lambda}/\pi$,
and the wave vector is commensurate. For $|\delta|>|\delta_c|$, domain
walls condense and the wave vector is incommensurate. In order to compute
the phase boundary, we need the magnetization for a given $\bar h_\perp$
and $\bar J_1$ close to the transition when $\alpha=1$. Within our chain-MFT we can compute
this by optimizing the mean-field energy. We find,
\begin{equation}  
|\Phi|^2_{\alpha=1}=\frac{h^{4a-2}_\perp}{36 G_4}\frac{ \bar J_1-\bar J^c_1}{\bar J_1\bar J^c_1}
\end{equation}
where $\bar J_1^c = h^{2a-2}_\perp/(6|G_2|)$. Now using this as input we
can calculate for $1-\alpha \ll 1$ the commensurate-incommensurate
phase boundary. We find at leading order in $(1-\alpha)$:
\begin{equation}
\bar J^{\rm CIT}_1 = \bar J^c_1+\mathcal{A}_{\rm CIT} \bar h_\perp^{2a-2} (1-\alpha)
\end{equation}
where $\mathcal{A}_{\rm CIT}=\frac{G_4 \pi}{-2G_2\sqrt{16G_4^2-6G_2G_6}}$ is a
pure number. 

\subsection{Ordering vector in the IC}
\label{sec:qstar_shift}

Once the $\bar h_i$ condense, initially at some incommensurate vector $q_x^*$, it is
clear that the ordering vector will change continuously in the {\bf IC}
phase as $\bar h_\perp$ is tuned. We can calculate the evolution of the ordering
vector from $q_x^*$ by performing a harmonic expansion valid close to the transition.
Approaching the {\bf IC} phase at high $\bar h_\perp$ from the {\bf PM} state, we
look for mean-field solutions including a higher harmonic, $\bar
h_i = \Phi_k e^{ik\cdot r_i} +\Phi_{3k} e^{i3k\cdot r_i} +{\rm c.c.}
$, which minimize the
energy, Eq.~(\ref{eq:var_free}). We can write the energy density as,
$e=\alpha_k|\Phi_k|^2+u_k|\Phi_k|^4 +
\alpha_{3k}|\Phi_{3k}|^2+\lambda_k (\Phi^3_k\Phi^*_{3k}+{\rm
  c.c.})$. Optimizing over $\Phi_k$ and $\Phi_{3k}$, we find that
minimum energy density is
$e=-\frac{\alpha_k^2}{4u_k}-\frac{\lambda_k^2}{\alpha_{3k}}\left
  (\frac{-\alpha_k}{2u_k}\right )^3$. If $\lambda_k=0$ then the
optimal $k=q_x^*$, but in practice there is a shift, that to leading
order in $(1-\alpha)$ and $\bar h_\perp - \bar h_\perp^c$ is given as,
\begin{equation}  
\delta q_x  = \frac{32(2a-2)^2}{9\sqrt{3}}(1-\alpha)\left(\frac{\bar h_\perp -
\bar h_\perp^c}{\bar h_\perp^c}\right)^2
\end{equation}
from which we conclude that the shift is positive for $(1-\alpha) \ll 1$.

\begin{figure}
  \includegraphics[width=5in]{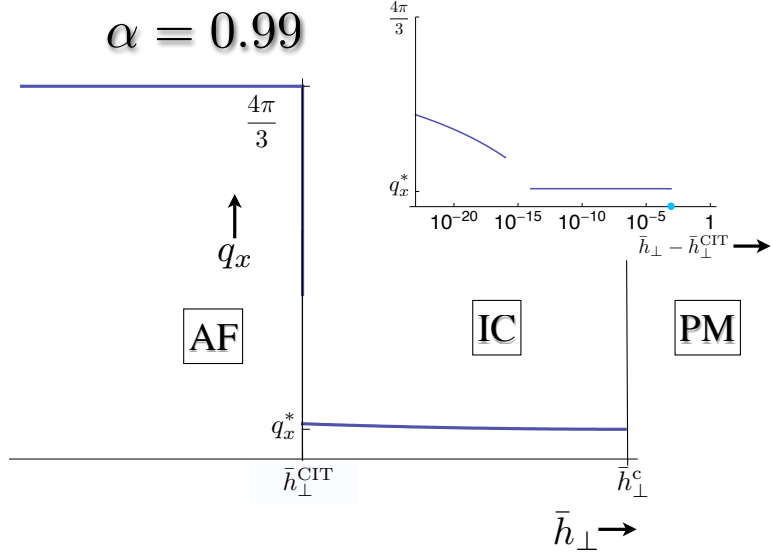}
  \caption{ \label{fig:qvec_inc} Primary ordering vector in the {\bf
      IC} phase. We have computed (in Section \ref{sec:qstar_shift}) the asymptotics expanding around
    the CIT and the transition from the {\bf PM} phase. The evolution
    of the $q_x$ is very sharp as shown. The inset shows the same
    data on a linear-log plot. The plot has been made for
    $\alpha=0.99$ and $J_1=0.1$ which gives, $q_x^*\approx4.17726$, $\bar
    h^{\rm c}_\perp\approx 1.05586$ and $\bar h^{\rm CIT}_\perp\approx 0.99842$.
}
 
\end{figure}

We now turn to the opposite edge of the {\bf IC} phase, when it
borders a commensurate phase, {\em i.e.} the {\bf AF} phase where the
ordering takes place at $(4\pi/3,0)$. Once we cross into the {\bf IC}
phase the finite density of domain walls shifts the ordering
vector. The shift 
is given by $\delta q_x = 2\pi n_d$, where $n_d$ is the linear
density of domain walls. In order to estimate $n_d$, we write an
expression for the
energy of a lattice of domain walls including an interaction energy between
domain wall pairs from the SG model (assuming $n_d w \ll 1$),
\begin{equation}
\label{eq:dwall_en}
\frac{E_{\rm sg}(n_d)}{A_{yz}} \approx \left(\frac{8 \sqrt{\kappa \lambda}}{n_{\rm sg}}
  -\frac{2 \pi \delta_x}{n_{\rm sg}}\right) n_d + U n_de^{-\frac{1}{wn_d}},
\end{equation}
where $U>0$ is the strength of repulsion between domain walls and
$w=\frac{1}{n_{\rm sg}} \sqrt{\frac{\kappa}{\lambda}}$ is the width of
domain walls and controls the range over which they interact. When
$\delta_x>\delta_c=4\sqrt{\kappa\lambda}/\pi$, there is a finite density of
domain walls, $n^*_d\sim\frac{1}{w{\rm
    ln}|\delta_c-\delta_x|}$. From this we can estimate the leading
singularity in the shift
in the ordering vector close to the CIT,
\begin{equation}
\delta q_x = \frac{-2\pi}{w\log [\bar h_\perp-\bar h^{\rm CIT}_\perp]}
\end{equation} 
where $w$ is the width of the domain wall at the CIT. From our
derived expressions Eq.~(\ref{eq:cg_cplg}) for the coupling constants
in the SG model, we can calculate the leading dependence of the domain
wall width at the CIT, we find,
$w=\frac{2\sqrt{3}}{\pi (1-\alpha)n_{\rm sg}}$. A plot of the two
asymptotes for the ordering vector is shown in Fig.~\ref{fig:qvec_inc}.

\begin{table}[t]
\begin{tabular}{||c||c|c|c||} 
\hline\hline
 & $n=1$ & $n=2$ & $n=3$  \\
 \hline\hline
$G_{2n}$ & $-1.84522$ & $8.33370$ & $-95.1689$ \\
\hline 
$\tilde G_n$ & $-1.35783$ & $-0.0489532$ & $0.0387529$\\
\hline 
$\tilde \Phi_{n-1}$ & $-1.19773$ & $-0.318819$ & $0.1108915$\\
\hline
\hline 
\end{tabular}
\caption{Leading coefficients of the energy expansion of the
  Ising field theory, appearing in Eq.~(\ref{eq:fz}), following Ref.~\onlinecite{fz}.}
\label{table:Gn}
\end{table}

\end{document}